\documentclass{elsart}
\usepackage{amssymb,amsfonts,amsmath,graphicx,subfigure}
\usepackage{graphics,graphicx,epsfig}
\begin{document}
\begin{frontmatter}
\title{Statistical analysis of the price index of Tehran Stock Exchange}

\author[a,b]{A. Rasoolizadeh}, \ead{$a\_rasooli@fardaorg.ir$}
\author[a]{R. Solgi\corauthref{e}}\ead{$r\_solgi@fardaorg.ir$}
\corauth[e]{Corresponding author. Tel.: +98-21-8950515-120; Fax:
+98-21-8980227; P.O.Box: 13595-369, Theran, Iran}

\address[a]{Quantitative Analysis Group, Tose-e-Farda Institute,\\
Tehran, Iran.}
\address[b]{Department of Economics, Allameh Tabatabaie University,\\
Tehran, Iran.}

\begin{abstract}
This paper presents a statistical analysis of Tehran Price Index
(TePIx) for the period of 1992 to 2004. The results present
asymmetric property of the return distribution which tends to the
right hand of the mean. Also the return distribution can be fitted
by a stable L\'{e}vy distribution and the tails are very fatter than
the gaussian distribution. We estimate the tail index of the TePIx
returns with two different methods and the results are consistent
with the previous studies on the stock markets. A strong
autocorrelation has been detected in the TePIx time series
representing a long memory of several trading days. We have also
applied a Zipf analysis on the TePIx data presenting strong
correlations between the TePIx daily fluctuations. We hope that this
paper be able to give a brief description about the statistical
behavior of financial data in Iran stock market.
\end{abstract}

\end{frontmatter}

\section{Introduction}
\label{Introduction} The large amount of available data and the
complexity of market structures has attracted a considerable
interest in recent years. The related researches has focused on
detailed statistical analysis of price fluctuations
\cite{Li,Mantegna1,Mantegna,Vandewalle1} and modeling markets as
complex
interactive systems \cite{Takayasu,Levy,Caldarelli}.\\
Tehran Stock Exchange opened in  1967. By the end of Iran's War and
the beginning of five year development plans in 1989, the market
observed a considerable growth (see Fig. 1-a), and now Tehran Stock
Exchange is the biggest and most active stock market in the middle
east area. In this paper Tehran Price Index (TePIx) is analyzed for
the period of 1992 to 2004 using the daily closing price index of
Tehran Stock Exchange excluding the intervals when the market was
closed.

\section{Distribution of the TePIx returns}

For the time series $P(t)$ which is TePIx on the day $t$, the return
$R(t)$ is defined as follows:

\begin{equation}
R(t)=\ln\frac{P(t+1)}{P(t)}\approx\frac{P(t+1)-P(t)}{P(t)}
\end{equation}

About a century ago Bachelier proposed the first model for the
return process \cite{Bachelier}. His model assumes a random walk
with Gaussian probability distribution function (PDF). But the large
changes in price which are very frequent in financial time series
\cite{Mantegna,Mandelbrot} and leads to fat tail
distributions, can not be modeled by a Gaussian process.\\
\begin{figure}[h]
\begin{center}
\includegraphics[width=14cm]{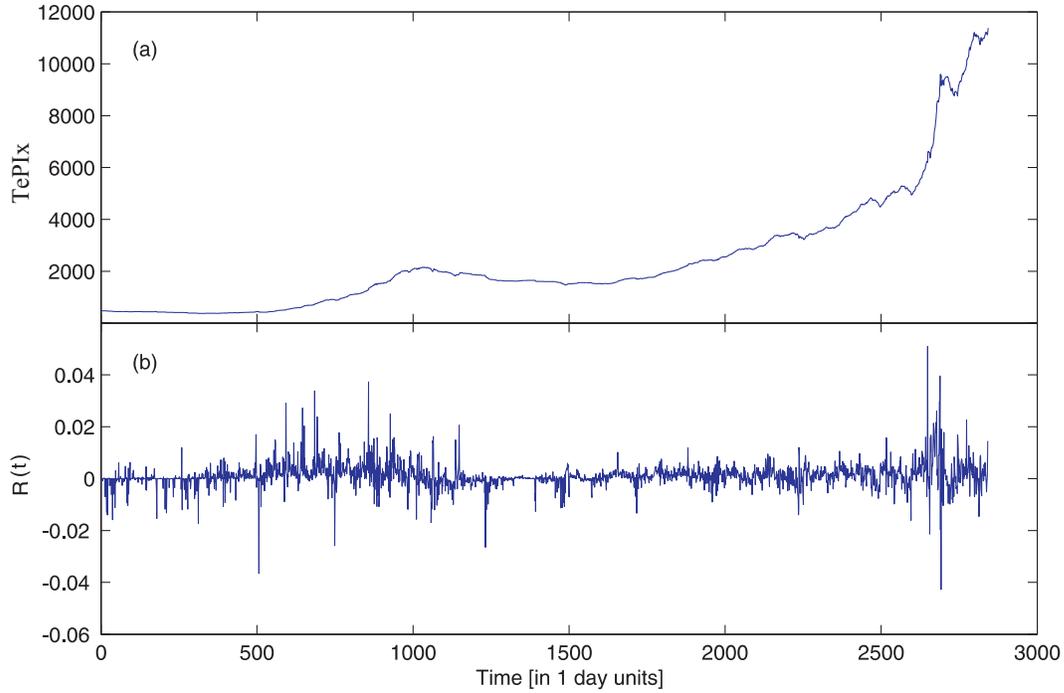}
\end{center}
\caption{TePIx values (a) and returns of the index (b) as a function
of time in 1 day units for the period of 1992 to 2004.}
\label{TePIx}
\end{figure}In the beginning of analyzing the distribution of TePIx returns (see
Fig. 1-b), mean, standard deviation, skewness, and kurtosis of the
return series are calculated (see Table \ref{mean}). The positive
value of skewness $\lambda_{3}=1.0619$, presents the asymmetric
property of the return distribution which tends to the right hand of
the mean. Indeed the large value of kurtosis $\kappa=20.827$ in
respect of Gaussian kurtosis ($\kappa=3$), shows that the tails of
the return
distribution are very fatter than the Gaussian ones.\\
\begin{table}[htb]
\begin{center}
\caption{\label{mean}Mean, standard deviation, skewness, and
kurtosis of the TePIx returns.}
\medskip
\begin{tabular}{ccccccccccccc}
\hline\hline $Mean$&$Std. Dev.$&$Skewness$&$Kurtosis$\\\hline
0.0011 & 0.0046 & 1.0619 & 20.827 \\
\hline\hline
\end{tabular}
\end{center}
\end{table}

\begin{figure}[h]
\begin{center}
\includegraphics[width=11cm]{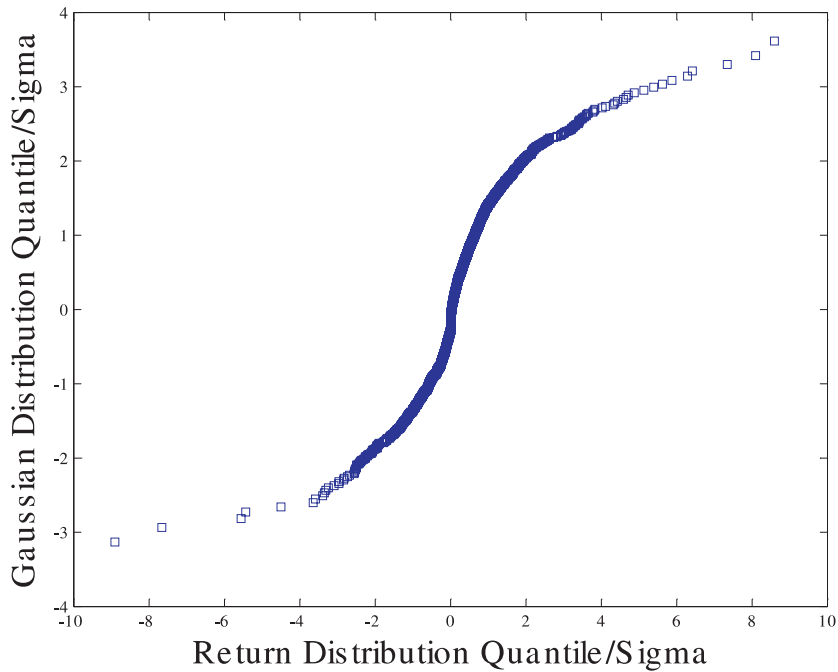}
\end{center}
\caption{Quantile-quantile plot of $R(t)$ distribution against
Gaussian PDF with the same mean and standard deviation.}
\label{quantile}
\end{figure}

\begin{figure}[h]
\begin{center}
\includegraphics[width=11cm]{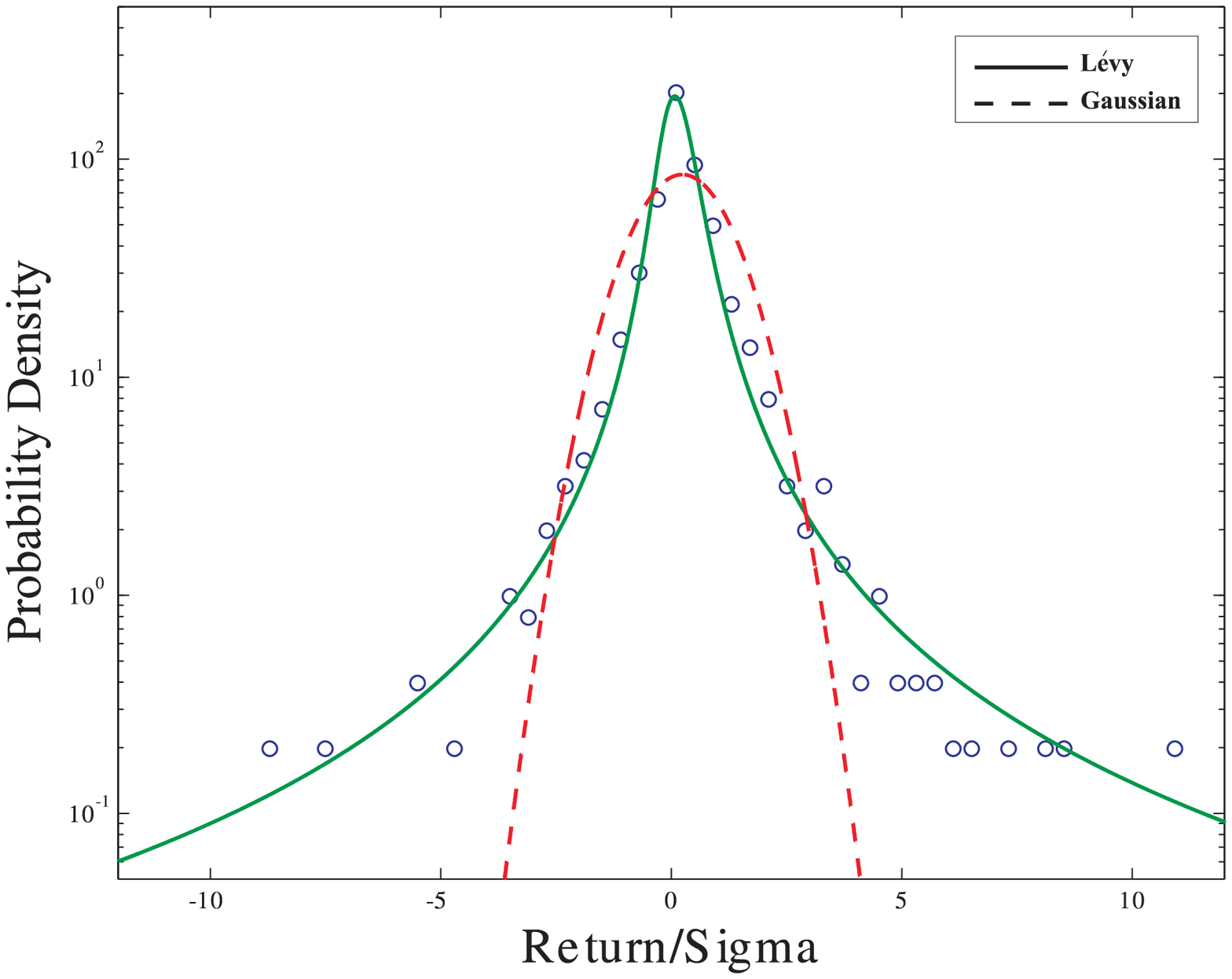}
\end{center}
\caption{Histogram of the daily returns of TePIx fitted by a stable
L\'{e}vy distribution.} \label{levy}
\end{figure}
For a better compression of the return distribution with a Gaussian
PDF, a quantile-quantile plot of $R(t)$ distribution against
Gaussian PDF with the same mean and standard deviation is depicted
in Fig. \ref{quantile}. If the PDF of returns was gaussian, all
points should have fallen on a straight line. It is seen that the
gaussianity is not a good approximation of this distribution and the
tails are much fatter than the gaussian distribution and therefor
displays the leptokurtic behavior of the returns. Also the histogram
of the daily returns of the TePIx is shown in Fig. \ref{levy}
(circles). It is obvious that the events which are 5 times larger
than the standard deviation of the returns (especially
in the right hand of average) is very frequent.\\
Also it is obvious that this distribution can be fitted by a stable
L\'{e}vy distribution \cite{Vries} (blue line). L\'{e}vy stable
distributions arise from the generalization of the central limit
theorem to a wider class of distributions. Consider the partial sum
$P_n\equiv\sum_{i=1}^{n}x_i$ of independent identically distributed
(i.i.d.) random variables $x_i$. If the $x_i$'s have finite second
moments, the central limit theorem holds and $P_n$ is distributed as
a Gaussian in the limit $n\rightarrow\infty$. If the random
variables $x_i$ are characterized by a distribution having
asymptotic power-law behavior:

\begin{equation}
P(x)\sim x^{-(1+\alpha)}
\end{equation}

where $\alpha<2$, then $P_n$ will converge to a L\'{e}vy stable
stochastic process of index $\alpha$ in the limit
$n\rightarrow\infty$. Except for special cases, such as the Cauchy
distribution $(\alpha=1)$ or the gaussian distribution $(\alpha=2)$,
L\'{e}vy distributions cannot be expressed in closed form. They are
often expressed in terms of their Fourier transforms or
characteristic functions, which we denote $\varphi(q)$, where $q$
denotes the Fourier transformed variable. The general form of a
characteristic function of a L\'{e}v stable distribution is:

\begin{equation}
\ln\varphi(q) = \left\{
\begin{array}{ll}
i\mu q-\gamma |q|^\alpha \left[1+i\beta \frac{q}{|q|}\tan\left(\frac{\pi}{2} \alpha \right)\right]&{\mathtt{~~~~}} [\alpha\neq1] \\
i\mu q-\gamma |q| \left[1+i\beta \frac{q}{|q|}\frac{2}{\pi} \ln|q|\right]&{\mathtt{~~~~}} [\alpha=1] \\
\end{array} \right. \label{Eq:levy}
\end{equation}

where $\alpha\in(0,2]$ is an index of stability also called the tail
index, $\beta\in[-1,1]$ is a skewness or asymmetry parameter,
$\gamma>0$ is a scale parameter, and $\mu\in\mathbb{R}$ is a
location parameter which is also called min.\\
The parameters of this fitted L\'{e}vy distribution are presented in
Table \ref{levypar}. Also a Gaussian PDF with the same mean and
standard deviation is plotted in the Fig. \ref{levypar}. It can be
seen that the tails of the real distribution (or the L\'{e}vy fitted
ones) are very fatter than the Gaussian tails.
\begin{table}[htb]
\begin{center}
\caption{\label{levypar}The parameters of the fitted L\'{e}vy
distribution.}
\medskip
\begin{tabular}{ccccccccccccc}
\hline\hline $\alpha$&$\beta$&$\gamma$&$\mu$\\\hline
1.213358 & 0.174998 & 0.0015315 & 0.000471761 \\
\hline\hline
\end{tabular}
\end{center}
\end{table}

\section{Tail index of the TePIx returns}
\subsection{Power law fit}
We analyze the asymptotic behavior of the cumulative distribution
function (CDF) of the TePIx returns too. It has been observed that
the right tail of CDF of returns can be fitted by a power law with
an exponent $\alpha_{R}=3.155\pm0.099$ in the
$\frac{R(t)}{\sigma}>3$ region, (see Fig. 4-b).\\
Also the left tail in the $\frac{R(t)}{\sigma}<-2$ region can be
fitted by a power law with an exponent $\alpha_{L}=3.022\pm0.118$,
(see Fig. 4-a).\\
Table \ref{powertail} includes the positive and
negative tails of the TePIx returns calculated with the power law
fitting method. These results are consistent with the previous
studies both on stock markets and foreign exchange markets
\cite{Vries,Gopikrishnan}.

\begin{figure}[h]
\begin{center}
\includegraphics[width=14cm]{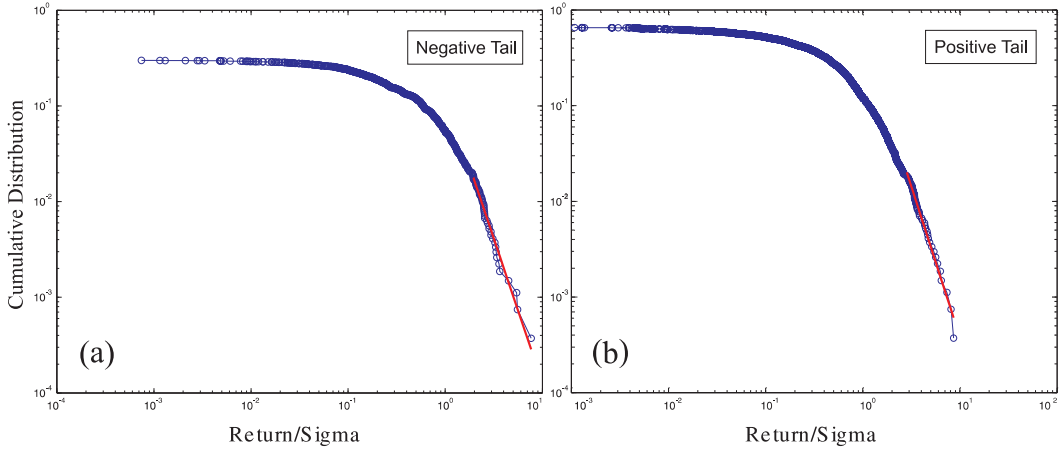}
\end{center}
\caption{Linear fit of the positive and negative tails of the
cumulative density function of the TePIx returns.} \label{PDFtails}
\end{figure}

\begin{table}[htb]
\begin{center}
\caption{\label{powertail}The tail index of the TePIx returns.}
\medskip
\begin{tabular}{ccc}
\hline\hline $Calculation\ Method$&$Positive\ tail$&$Negative\
tail$\\\hline
$Power law\ fit $ & $3.155\pm0.099$ & $3.022\pm0.118$ \\
\hline\hline
\end{tabular}
\end{center}
\end{table}

\subsection{Hill estimator method}
\begin{figure}[h]
\begin{center}
\includegraphics[width=14cm]{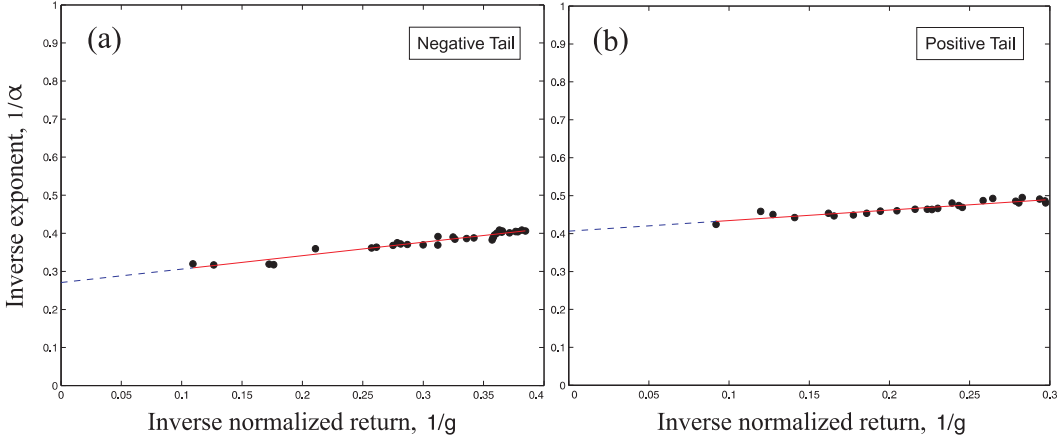}
\end{center}
\caption{The positive and negative tails of the cumulative density
function calculated with the hill estimator method.}
\label{hilltails}
\end{figure}

We have also used the Hill estimator method to obtain a more
accurate estimation of asymptotic behavior of the cumulative density
function \cite{Hill,stanley}, (see Fig. \ref{hilltails}). The basic
idea is to calculate the inverse of the local logarithmic slope
$\zeta$ of the cumulative distribution $P(g>x)$:

\begin{equation}
\zeta\equiv-{\left(\frac{d\ln P}{d\ln x}\right)}^{-1}
\end{equation}

We then estimate the inverse asymptotic slope $1/\alpha$ by
extrapolating $\zeta$ as $(1/x)\rightarrow0$. The descending sorted
normalized returns is denoted $g_{k}$, where $k=1,...,N$ and $N$ is
the total number of events. Then the inverse local slope of
$\zeta(g)$  can be written as:

\begin{equation}
\zeta(g_k)=\frac{\ln(g_{k+1}/g_k)}{\ln(P(g_{k+1})/P(g_k))}
\end{equation}

The above expression can be well approximated for large $k$ as:

\begin{equation}
\zeta(g_k)=k(\ln(g_{k+1})-\ln(g_k))
\end{equation}

The inverse local slopes is obtained through the above equation.
Then an average of the inverse slopes is computed over $m$ points:

\begin{equation}
\langle\zeta\rangle=\frac{1}{m}\sum_{k=1}^m \zeta(g_k)
\end{equation}

where the choice of the averaging window length $m$ varies depending
on the number of available events $N$. We plot the locally averaged
inverse slope $\langle\zeta\rangle$ as a function of the inverse
normalized returns $1/g$. Then the $\zeta$ is extrapolated as a
function of $1/g$ to 0. This procedure yields the inverse asymptotic
slope $1/\alpha$. Table \ref{hilltail} includes the positive and
negative tails of the TePIx returns calculated with the Hill
estimator method.

\begin{table}[htb]
\begin{center}
\caption{\label{hilltail}The tail index of the TePIx returns.}
\medskip
\begin{tabular}{ccc}
\hline\hline $Calculation\ Method$&$Positive\ tail$&$Negative\
tail$\\\hline
$Hill\ estimator $ & $2.4639\pm0.095$ & $3.708\pm0.2071$ \\
\hline\hline
\end{tabular}
\end{center}
\end{table}

\section{Correlation structure of Tehran Stock Exchange}
\subsection{Autocorrelation function of the TePIx returns}
Autocorrelation is a commonly used method for checking randomness in
a data set. The following equation is the autocorrelation function
of a time series, in which $l$ denotes non negative varying time
lags of the data set:

\begin{equation}
AC(R_{t},l)=\frac{<R_{(t+l)}R_{t}>}{<R_{t}^2>}
\end{equation}

If the time series is random, the autocorrelations should be near
zero for any and all time lag separations. If it is not random, then
one or more of the autocorrelations will be significantly non zero
\cite{Box}.

\begin{figure}[h]
\begin{center}
\includegraphics[width=11cm]{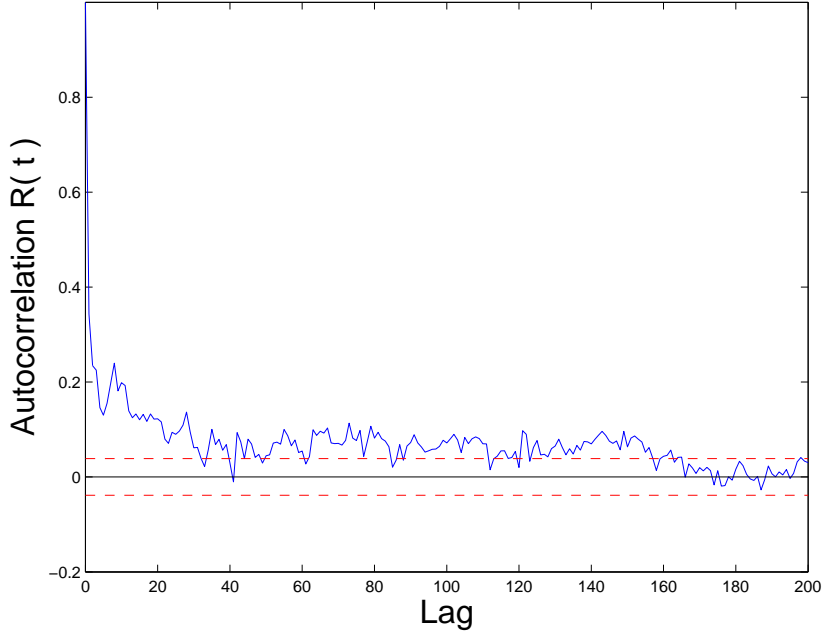}
\end{center}
\caption{Autocorrelation function of the return of the TePIx time
series.} \label{ac200}
\end{figure}

\begin{figure}[h]
\begin{center}
\includegraphics[width=11cm]{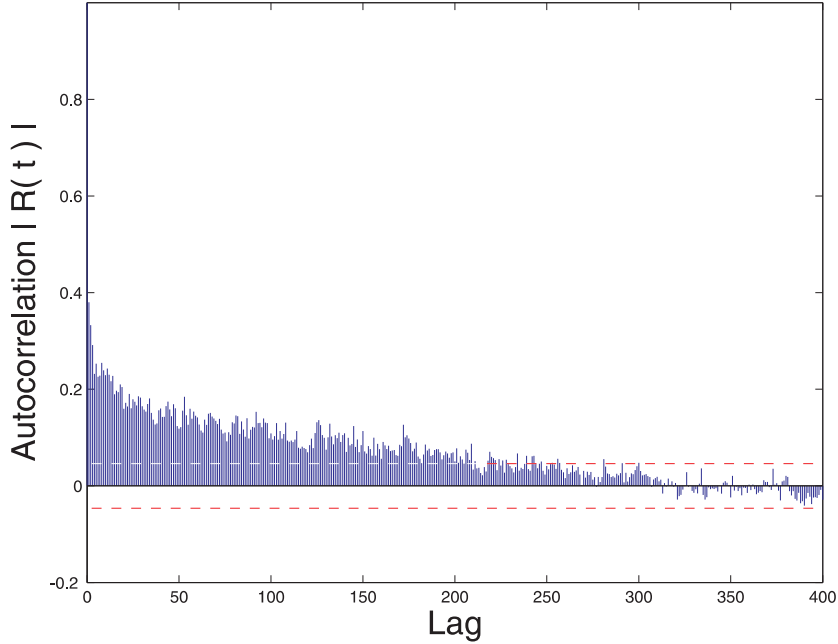}
\end{center}
\caption{Autocorrelation function of the modulus TePIx returns.}
\label{ac400}
\end{figure}

In Fig. \ref{ac200} the dotted lines are the 95\% confidence band
and as it is seen the autocorrelation decays slowly and therefor the
TePIx has a memory of several trading days. There is an evidence of
considerable positive autocorrelation for the values of
$l\leqslant30$, after which the autocorrelations are at the level of
noise. The autocorrelation function of the modulus time series (see
Fig. \ref{ac400}), that is the absolute returns without regarding
the sign, displays a very long term memory. This indicates that the
volatility is clustered in time.

\subsection{Persistence analysis of the TePIx time series}
A common way for persistence analysis is to compute the histogram
for the step length of monotonous index changes. In order to do so,
we build a new series where the trading days will be distributed in
the clusters of different sizes characterized by $l^+$ and $l^-$,
expressing the monotonous increase or decrease of the Index. $l^+$
and $l^-$ respectively denote the number of days in which the index
increases or decreases monotonously. In an unbiased sequence when
there is no correlation in the market, the number of observations of
$l^\pm$ in $N$ continuous days equals:

\begin{equation}
P(l^+ )=(N-l^+ +1)P_{u}^{l^+}\label{unbiased1}
\end{equation}

\begin{equation}
P(l^- )=(N-l^- +1)P_{d}^{l^-}\label{unbiased2}
\end{equation}

where letters $u$ and $d$, represent the up and down fluctuations of
the series, respectively. In an unbiased random sequence it is
expected that the frequencies of $u$ and $d$ are equal, in other
words we call a sequence unbiased if $P_{u}$ (the fraction of $u$'s)
is equal to $P_{d}$ (fraction of $d$'s). However the situation is
different in a biased case (e.g. TePIx), in which
$P_{u}=\frac{1}{2}+\varepsilon$ and $P_{d}=\frac{1}{2}-\varepsilon$
where $\varepsilon\in[0,\frac{1}{2}]$:

\begin{equation}
P(l^\pm )=(N-l^\pm +1)(\frac{1}{2}
+\varepsilon)^{l^\pm}\label{biased1}
\end{equation}

%\begin{equation}
%P(l^- )=(N-l^- +1)(\frac{1}{2} +\varepsilon)^{l^-}\label{biased2}
%\end{equation}

If $N$ is considerably greater than $l^\pm$ ($N\gg l^\pm$), then
$P(l^\pm )$ vs. $l^\pm$ expresses an exponential behavior in the
form of:

\begin{equation}
P(l)=\alpha \exp (-\beta|l|)
\end{equation}

\begin{figure}[h]
\begin{center}
\includegraphics[width=11cm]{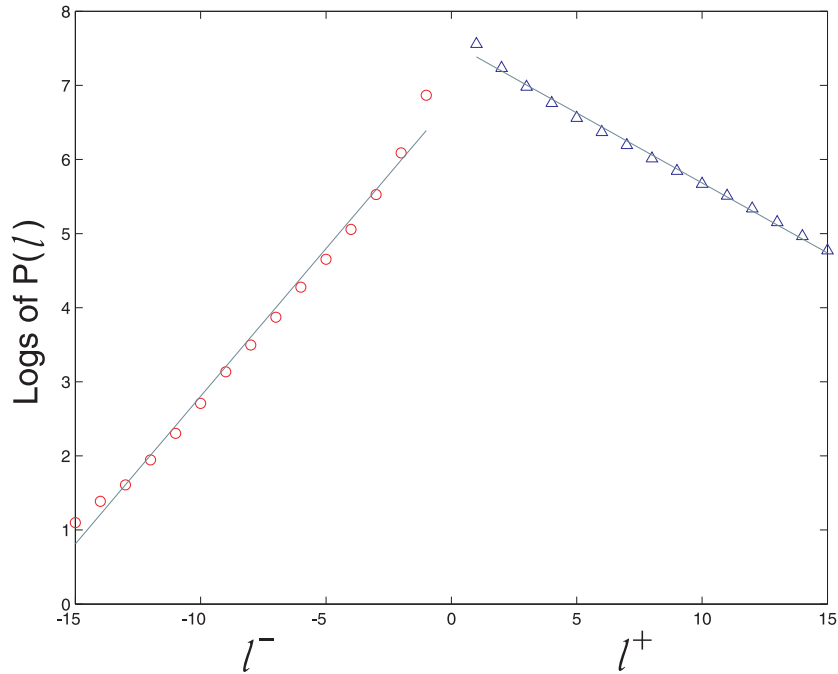}
\end{center}
\caption{Histogram for the step lengths of monotonous index
changes.} \label{persistance}
\end{figure}

In other words, the logarithm of $P(l^+ )$ vs. $l^+$ must be a line
with a $\ln[P(l^+ )]$ slope, and the logarithm of $P(l^- )$ vs.
$l^-$ must be a line with a $\ln[P(l^- )]$ slope. The lower and
greater values of the slopes are an indication of persistence
and anti-persistence in the time series respectively.\\
As it is seen in Fig. \ref{persistance}, the histogram of monotonous
index changes is well fitted with an exponential distribution with
the estimated parameters presented in table \ref{tpersistance}.

\begin{table}[htb]
\begin{center}
\caption{\label{tpersistance}Parameters of the fitted power law on
the monotonous index changes.}
\medskip
\begin{tabular}{ccc}
\hline\hline $Powerlaw\ fitting$&$Negative\ trend$&$Positive\
trend$\\\hline
$Estimated\ \beta$ & $0.39837\pm0.02439$ & $0.18882\pm0.00855$ \\
\hline\hline
\end{tabular}
\end{center}
\end{table}

As it is seen in the Fig. 1-a there is an intensive drift in the
TePIx time series with the following probabilities of increase and
decrease in the index:

\ $~$ \hspace{1cm}$P_{down} =0.31403$ \hspace{1cm}$P_{constant}
=0.05888$ \hspace{1cm}$P_{up} =0.62709$

In a random walk with a bias similar to TePIx and in the lack of
correlation, the expected parameters ($\beta_d
=\ln(P_{down})=1.1583,\ \beta_u =\ln(P_{up})=0.46667$) would be much
greater than the fitted parameters, thus there is a very strong
persistence in Tehran Stock Exchange.\\
The distribution (See Fig. \ref{persistance}) is completely
asymmetric and the probability of positive changes of length $l$ is
completely more than the probability of a negative run of the same
length which consists with the intensive drift in the index series.
This model implies that the probability of the next step continuing
the increasing trend can be estimated as $P(l^+ |~ l^+
-1)=\exp(-0.18882)=0.82793$ and the probability of the next step
continuing the decreasing trend can be estimated as $P(l^- |~ l^-
-1)=\exp(-0.39837)=0.67141$. Both  of them are much greater than the
probability that would be obtained from a biased random walk
(Equation \ref{biased1}) similar to TePIx, which is $P(l^+ |~ l^+
-1)=P_{up} =0.62709$ and $P(l^- |~ l^- -1)=P_{down} =0.31403$
respectively.

\section{Zipf analysis of Tehran Stock Exchange}

Zipf law is an interesting feature of natural languages. According
to Zipf law, If all the words in a text are sorted based on their
frequency of appearance in a descending order, a power law with an
exponent $\zeta$ will b appeared \cite{Zipf}:

\begin{equation}
    f\propto R^{-\zeta}
\end{equation}

where $f$ is the frequency of appearance of a word, and $R$ is its
rank in the sorted list of the words, with $\zeta\approx1$ for all
languages that have been studied. The origin of this scaling is the
hierarchical structure and existence of long range correlations
between words in a text. Recently Zipf analysis has been applied to
study the various complex systems in different contexts.

\begin{figure}[h]
\begin{center}
\includegraphics[width=11cm]{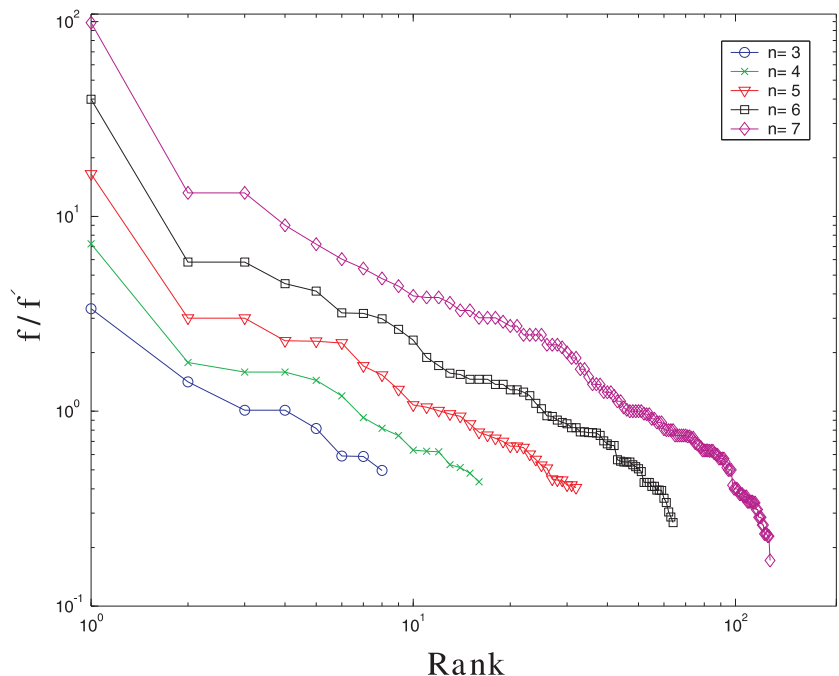}
\end{center}
\caption{Zipf diagram of the TePIx data for $n=3, 4, 5, 6$ and $7$.}
\label{zipf1}
\end{figure}

\begin{table}[htb]
\begin{center}
\caption{\label{tbzipf} Apparent frequencies $f$, effective
frequencies $f/f^{'}$, error bars of effective frequencies
$\delta_{f/f^{'}}$ and rank of words $R$, for the words of size 4.}
\medskip
\begin{tabular}{cccccc}
\hline\hline
$Word$&$f$&$f^{'}$&$f/f^{'}$&$\delta_{f/f^{'}}$&$Rank$\\\hline
uuuu & 0.069867 & 0.0096533 & 7.2377 & 0.20011 & 1 \\
uuud & 0.033568 & 0.021144 & 1.5876 & 0.13443 & 2 \\
uudu & 0.025371 & 0.021144 & 1.1999 & 0.13443 & 3 \\
uudd & 0.042935 & 0.046311 & 0.92711 & 0.089655 & 6 \\
uduu & 0.030445 & 0.021144 & 1.4399 & 0.13443 & 4 \\
udud & 0.02459 & 0.046311 & 0.53098 & 0.089655 & 7 \\
uddu & 0.02381 & 0.046311 & 0.51413 & 0.089655 & 8 \\
uddd & 0.063232 & 0.10143 & 0.62338 & 0.058802 & 12 \\
duuu & 0.033568 & 0.021144 & 1.5876 & 0.13443 & 5 \\
duud & 0.034738 & 0.046311 & 0.75012 & 0.089655 & 9 \\
dudu & 0.029274 & 0.046311 & 0.63212 & 0.089655 & 10 \\
dudd & 0.044106 & 0.10143 & 0.43483 & 0.058802 & 13 \\
dduu & 0.037861 & 0.046311 & 0.81754 & 0.089655 & 11 \\
ddud & 0.04879 & 0.10143 & 0.4810 & 0.058802 & 14 \\
dddu & 0.062842 & 0.10143 & 0.61953 & 0.058802 & 15 \\
dddd & 0.3950 & 0.22217 & 1.7779 & 0.036967 & 16 \\
\hline\hline
\end{tabular}
\end{center}
\end{table}

As the first step, to study the TePIx signal it should be translated
to a sequence of letters in an alphabet. For this purpose a binary
alphabet $\{u,d\}$ is considered, which it's letters $u$ and $d$,
represent the up and down fluctuations of the TePIx, respectively.
For a given $n$, there is $2^{n}$ word in this alphabet. In an
unbiased random sequence it is expected that these frequencies are
equal. We call a sequence unbiased if $p_{u}$ (the fraction of
$u$'s) is equal to $p_{d}$ (fraction of $d$'s). In this case the
Zipf plot is a horizontal line. However the situation is different
in a biased case: assume that $p_{u}=\frac{1}{2}+\varepsilon$ and
$p_{d}=\frac{1}{2}-\varepsilon$ where
$\varepsilon\in[0,\frac{1}{2}]$. In this case the frequency of any
$C^{k}_{n}$ words that include(s) exactly k $u$'s and n-k $d$'s is
proportional to $p_{u}^{k}p_{d}^{n-k}$. Then the Zipf plot
represents a non-zero slope which is approximately equal to
\cite{Vandewalle}:

\begin{equation}
    \zeta\approx-\frac{\ln(\frac{1}{2}-\varepsilon)/\ln(\frac{1}{2}+\varepsilon)}{\ln n}
\end{equation}

It should be noted that some small bias may cause large Zipf
exponents even for large values of n. This non-zero Zipf exponent is
due to the existence of a bias, not due to the existence of
correlations. To avoid this problem in the Zipf analysis of a
financial sequence, instead of the apparent frequencies of the words
$f$, the effective frequencies of them $f/f^{'}$ is applied, where
$f{'}$ is the expected frequency of a random sequence with the same
bias. In this manner, random biased sequences present a zero Zipf
exponent, too. If the log-log plot of $f/f^{'}$ vs. $R$ reveals
some negative slopes, it means that there are some non trivial correlations in the sequence.\\
The evolution of TePIx in the mentioned period is shown in the Fig.
1-a. As it can be seen there is a positive trend in this evolution.
After translation of this signal to a string in $\{u,d\}$ alphabet,
$p_{u}$, $p_{d}$ and $\varepsilon$ can be calculated. This signal is
biased and we have $\varepsilon=0.1865$. Then the Zipf analysis has
been done on this sequence. Zipf plot for $n=3, 4, 5, 6$ and $7$ are
depicted in Fig. \ref{zipf1}. We see that although the effective
frequencies $f/f^{'}$ is applied in the vertical axis, but a
negative Zipf exponent which is approximately equal to $0.9$ is
observed. It means that there are strong correlations between the
TePIx daily fluctuations. Also for the $n=4$, apparent frequencies
$f$, effective frequencies $f/f^{'}$, error bars of effective
frequencies $\delta_{f/f^{'}}$ and rank of words $R$, can be seen in
Table \ref{tbzipf}.

\section{Conclusions}
This paper presented a statistical analysis of Tehran Price Index
(TePIx) for the period of 1992 to 2004. The positive value of
skewness $\lambda_{3}=1.0619$, presents the asymmetric property of
the return distribution which is skewed to right and the large value
of kurtosis $\kappa=20.827$ in respect of Gaussian kurtosis
($\kappa=3$), shows that the tails of the return distribution are
very fatter than the Gaussian tails. Also it is demonstrated that
the return distribution can be fitted by a stable L\'{e}vy
distribution.
\\
We examined the tail behavior of the return distribution with two
different methods and the results are consistent with the previous
studies on the stock markets. Also there is an evidence of
considerable positive autocorrelation for the values of
$l\leqslant30$, after which the autocorrelations are at the level of
noise. In the last section, a Zipf analysis is applied on the TePIx
data and the results present strong correlations between the TePIx
daily fluctuations.

\textbf{Acknowledgments}

We would like to thank Tose-e-Farda Institute for the financial
support of the research presented in this paper.


\begin{thebibliography}{99}

\bibitem{Li} W. Li, Int. J. Bifurcat. \emph{Chaos} 1 (1991) 583.
\bibitem{Mantegna1} R.N. Mantegna, \emph{Physica A} 179 (1991) 232.
\bibitem{Mantegna} R. N. Mantegna, H. E. Stanley, \emph{Nature} 376, 46 (1995).
\bibitem{Vandewalle1} N. Vandewalle, M. Ausloos, \emph{Physica A} 246 (1997) 454.
\bibitem{Takayasu} H. Takayasu, H. Miura, T. Hirabayashi, K. Hamada, \emph{Physica A} 184 (1992) 127.
\bibitem{Levy} M. Levy, H. Levy, S. Solomon, \emph{J. Phys. I} 5 (1995) 1087.
\bibitem{Caldarelli} G. Caldarelli, M. Marsili, Y.C. Zhang, \emph{Europhys. Lett.} 40 (1997) 479.
\bibitem{Bachelier} L. Bachelier, \emph{Ann. Sci. Ecole Norm. Suppl.} 3, 21 (1990).
\bibitem{Mandelbrot} B. B. Mandelbrot, \emph{J. Business} 36, 294 (1963).
\bibitem{Vries} C. G. de Vries, \emph{in the Handbook of International
Macroeconomics}, edited by F. van der Ploeg, (Blackwell, Oxford,
1994).
\bibitem{Gopikrishnan} P. Gopikrishnan, V. Plerou, L. A. Nunes
Amaral, M. Meyer, H. E. Stanley, \emph{Phys. Rev. E} 60, (1999),
5305-5316.
\bibitem{Hill} B. M. Hill, Ann. Stat. 3, 1163 (1975).
\bibitem{stanley} H. E. Stanley, Ph.D. thesis, Harvard University, (1967).
\bibitem{Box} G. E. P. Box, and G. Jenkins, \emph{Time Series Analysis:
Forecasting and Control}, Holden-Day, (1976)
\bibitem{Zipf} G. K. Zipf, \emph{Human Behavior and the Principle of Least
Effort}, (Addisson-Wesley, Cambridge, MA, 1949).
\bibitem{Vandewalle} N. Vandewalle, M. Ausloos, \emph{Physica A} 268 (1999)
240-249.

\end{thebibliography}
\end{document}